%% file: main.tex
\documentclass[sigconf,natbib=true]{acmart}
\AtBeginDocument{%
  \providecommand\BibTeX{{%
    \normalfont B\kern-0.5em{\scshape i\kern-0.25em b}\kern-0.8em\TeX}}}

\usepackage{booktabs} 

\usepackage[english]{babel}
\usepackage{moresize}
\usepackage{amsmath}
\usepackage{algorithmic} 
\usepackage{balance}
\usepackage{comment}
\usepackage{paralist}
\usepackage{bm}
\usepackage{pgfplots}
\pgfplotsset{compat=1.17}
\usetikzlibrary{pgfplots.dateplot}

\usepackage{flushend}
\usepackage[english]{babel}
\usepackage[latin1]{inputenc}
\usepackage{mathrsfs}
\usepackage{graphicx}

\usepackage{amssymb}
\usepackage{amsfonts}
\usepackage{url}
\usepackage{longtable}
\usepackage{rotating}
\usepackage{multirow}
\usepackage{mathrsfs}
\usepackage{subfigure}
\usepackage{enumitem}
\usepackage[linesnumbered,algoruled,boxed,lined]{algorithm2e}
\usepackage{adjustbox}
\usepackage{hyperref}
\usepackage{amssymb}
\usepackage{amsfonts}
\usepackage{pifont}
\DeclareMathOperator*{\argmax}{arg\,max}

\usepackage{pgfplots}
\usetikzlibrary{pgfplots.dateplot}
\usepackage{graphicx}
\usepackage{caption}
\usepackage{threeparttable}
\usepackage{enumitem}
\usepackage{nccmath}

\usepackage{diagbox}

\usepackage{filecontents}
\usepackage{tabularx}

\newcommand{\hide}[1]{} 

\newcommand{\ie}{\textit{i}.\textit{e}.}
\newcommand{\eg}{\textit{e}.\textit{g}.} 
\newcommand{\wrt}{\textit{w}.\textit{r}.\textit{t}}

\def\model{LLMRec}

\setcopyright{acmlicensed}
\copyrightyear{2024}
\acmYear{2024}
\acmDOI{10.1145/3616855.3635853}

\acmConference[WSDM '24]{Proceedings of the 17th ACM International Conference on Web Search and Data Mining}{March 4--8, 2024}{Merida, Mexico}
\acmBooktitle{Proceedings of the 17th ACM International Conference on Web Search and Data Mining (WSDM '24), March 4--8, 2024, Merida, Mexico}

\acmPrice{15.00}
\acmISBN{979-8-4007-0371-3/24/03}

\ccsdesc[500]{Information systems~Recommender systems}

\begin{document}

\title{ LLMRec: Large Language Models with Graph Augmentation \\ for Recommendation }  


\fancyhead{}

\author{Wei Wei}
\affiliation{%
  \institution{University of Hong Kong}
  \city{}
  \country{}}
\email{weiweics@connect.hku.hk}

\author{Xubin Ren}
\affiliation{%
  \institution{University of Hong Kong}
  \city{}
  \country{}}
\email{xubinrencs@gmail.com}

\author{Jiabin Tang}
\affiliation{%
  \institution{University of Hong Kong}
  \city{}
  \country{}}
\email{jiabintang77@gmail.com}

\author{Qinyong Wang}
\affiliation{%
  \institution{Baidu Inc.}
  \city{}
  \country{}}
\email{wangqinyong@baidu.com}

\author{Lixin Su}
\affiliation{%
  \institution{Baidu Inc.}
  \city{}
  \country{}}
\email{sulixinict@gmail.com}

\author{Suqi Cheng}
\affiliation{%
  \institution{Baidu Inc.}
  \city{}
  \country{}}
\email{chengsuqi@gmail.com}

\author{Junfeng Wang}
\affiliation{%
  \institution{Baidu Inc.}
  \city{}
  \country{}}
\email{wangjunfeng@baidu.com}

\author{Dawei Yin}
\affiliation{%
  \institution{Baidu Inc.}
  \city{}
  \country{}}
\email{yindawei@acm.org}

\author{Chao Huang}
\authornote{Chao Huang is the corresponding author.}
\affiliation{%
  \institution{University of Hong Kong}
  \city{}
  \country{}}
\email{chaohuang75@gmail.com}

\begin{abstract}

The problem of data sparsity has long been a challenge in recommendation systems, and previous studies have attempted to address this issue by incorporating side information. However, this approach often introduces side effects such as noise, availability issues, and low data quality, which in turn hinder the accurate modeling of user preferences and adversely impact recommendation performance. In light of the recent advancements in large language models (LLMs), which possess extensive knowledge bases and strong reasoning capabilities, we propose a novel framework called \underline{\model} that enhances recommender systems by employing three simple yet effective LLM-based graph augmentation strategies. Our approach leverages the rich content available within online platforms (e.g., Netflix, MovieLens) to augment the interaction graph in three ways: (i) reinforcing user-item interaction egde, (ii) enhancing the understanding of item node attributes, and (iii) conducting user node profiling, intuitively from the natural language perspective. By employing these strategies, we address the challenges posed by sparse implicit feedback and low-quality side information in recommenders. Besides, to ensure the quality of the augmentation, we develop a denoised data robustification mechanism that includes techniques of noisy implicit feedback pruning and MAE-based feature enhancement that help refine the augmented data and improve its reliability. Furthermore, we provide theoretical analysis to support the effectiveness of \model\ and clarify the benefits of our method in facilitating model optimization. Experimental results on benchmark datasets demonstrate the superiority of our LLM-based augmentation approach over state-of-the-art techniques. 
To ensure reproducibility, we have made our code and augmented data publicly available at: \url{ https://github.com/HKUDS/LLMRec.git }. 
\end{abstract}

\keywords{Large Language Models, Graph Learning, Data Augmentation, Content-based Recommendation, Multi-modal Recommendation, Collaborative Filtering, Data Sparsity, Bias in Recommender System}


\maketitle
\input{intro}

\input{pre}

\input{solution}

\input{eval}

\input{related}
\input{conclusion}

\clearpage
\balance
\bibliographystyle{ACM-Reference-Format}
\bibliography{sample-base}

\clearpage

\end{document}

%% file: intro.tex
\section{Introduction}
\label{sec:intro}


Recommender systems play a crucial role in mitigating information overload by providing online users with relevant content~\cite{meng2023hierarchical,wei2022contrastive}. 
To achieve this, an effective recommender needs to have a precise understanding of user preferences, which is not limited to analyzing historical interaction patterns but also extends to incorporating rich side information associated with users and items~\cite{zou2022multi}.

In modern recommender systems, such as Netflix, the side information available exhibits heterogeneity, including item attributes \citep{ying2023camus}, user-generated content \citep{fan2019graph,meng2023coarse}, and multi-modal features \cite{yi2022multi} encompassing both textual and visual aspects. This diverse content offer distinct ways to characterize user preferences. By leveraging such side information, models can obtain informative representations to personalize recommendations. However, despite significant progress, these methods often face challenges related to data scarcity and issues associated with handling side information.
\\\vspace{-0.12in}

\noindent \textbf{Sparse Implicit Feedback Signals}. 
Data sparsity and the cold-start problem hinder collaborative preference capturing \cite{wei2021contrastive}.
While many efforts (\eg, NGCF~\cite{wang2019neural},  LightCGN~\citep{he2020lightgcn}) tried powerful graph neural networks(GNNs) in collaborative filtering(CF), they face limits due to insufficient supervised signals. Some studies \cite{ren2023sslrec} used contrastive learning to add self-supervised signals 
(\eg, SGL~\cite{wu2021self}, SimGCL~\cite{yu2022graph}).
However, considering that real-world online platforms (\eg, Netflix, MovieLens) derive benefits from modal content, recent approaches, unlike general CF, are dedicated to incorporating side information as auxiliary for recommenders. 
For example, 
MMGCN \cite{wei2019mmgcn} and GRCN \cite{wei2020graph} incorporate item-end content into GNNs to discover high-order content-aware relationships.  
LATTICE \cite{zhang2021mining} leverages auxiliary content to conduct data augmentation by establishing i-i relationships.
Recent efforts (\eg, MMSSL \cite{wei2023multi}, MICRO \cite{zhang2022latent}) address sparsity by introducing self-supervised tasks that maximize the mutual information between multiple content-augmented views. 
However, strategies for addressing data sparsity in recommender systems, especially in multi-modal content, can sometimes be limited. This is because the complexity and lack of side information relevance to CF can introduce distortions in the underlying patterns \cite{wei2020graph}. Therefore, it becomes crucial to ensure the accurate capture of realistic user preferences when incorporating side information in CF, in order to avoid suboptimal results. \\\vspace{-0.12in}


\noindent \textbf{Data Quality Issues of Side Information}. 
Recommender systems that incorporate side information often encounter significant issues that can negatively impact their performance.
i) \textbf{Data Noise} is an important limitation faced by recommender systems utilizing side information is the issue of data noise\cite{tian2022learning}, where attributes or features may lack direct relevance to user preferences. For instance, in a micro video recommender, the inclusion of irrelevant textual titles that fail to capture the key aspects of the video's content introduces noise, adversely affecting representation learning.
The inclusion of such invalid information confuse the model and lead to biased or inaccurate recommendations. 
ii) \textbf{Data heterogeneity}\citep{chen2023heterogeneous} arises from the integration of different types of side information, each with its own unique characteristics, structures, and representations. 
Ignoring this heterogeneity leads to skewed distributions \citep{ying2023camus,meng2023parallel}. Bridging  heterogeneous gap is crucial for successfully incorporating side information uniformly. iii) \textbf{Data incompleteness} \citep{ko2022survey,liang2023learn} 
occurs when side information lacks certain attributes or features. For instance, privacy concerns\cite{zhang2023lightfr} may make it difficult to collect sufficient user profiles to learn their interests. Additionally, items may have incomplete textual descriptions or missing key attributes. This incompleteness impairs the model's ability to fully capture the unique characteristics of users and items, thereby affecting the accuracy of recommendations.

Having gained insight into data sparsity and low-quality encountered by modern recommenders with auxiliary content, 
this work endeavors to overcome these challenges through explicit augment potential user-item interactive edges as well as enhances user/item node side information (\eg, language, genre). 
Inspired by the impressive natural language understanding ability of large language models (LLMs), we utilize LLMs to augment the interaction graph.
\emph{Firstly}, \model\ embraces the shift from an ID-based recommendation framework to a modality-based paradigm \citep{li2023text, yuan2023go}. It leverages large language models (LLMs) to predict user-item interactions from a natural language perspective. Unlike previous approaches that rely solely on IDs, \model\ recognizes that valuable item-related details are often overlooked in datasets \citep{li2023gpt4rec}. Natural language representations provide a more intuitive reflection of user preferences compared to indirect ID embeddings. By incorporating LLMs, \model\ captures the richness and context of natural language, enhancing the accuracy and effectiveness of recommendations.
\emph{Secondly}, to elaborate further, the low-quality and incomplete side information is enhanced by leveraging the extensive knowledge of LLMs, which brings two advantages:
i) 
LLMs are trained on vast real-world knowledge, allowing them to understand user preferences and provide valuable completion information, even for privacy-constrained user profiles.
ii) The comprehensive word library of LLMs unifies embeddings in a single vector space, bridging the gap between heterogeneous features and facilitating encoder computations. This integration prevents the dispersion of features across separate vector spaces and provide more accurate results.

Enabling LLMs as effective data augmentors for recommenders poses several technical challenges that need to be addressed:  \\\vspace{-0.22in}
\begin{itemize}[leftmargin=*]
\item \textbf{C1:} How to enable LLMs to reason over user-item interaction patterns by explicitly augmenting implicit feedback signals? \\\vspace{-0.12in}
\item \textbf{C2:} 
How to ensure the reliability of the LLM-augmented content to avoid introducing noise that could compromise the results?  \\\vspace{-0.22in}  
\end{itemize} 
The potential of LLM-based augmentation to enhance recommenders by addressing sparsity and improving incomplete side information is undeniable. However, effectively implementing this approach requires addressing the aforementioned challenges. Hence, we have designed a novel framework \underline{LLMRec} to tackle these challenges. \\\vspace{-0.12in}

\noindent \textbf{Solution.}
Our objective is to address the issue of sparse implicit feedback signals derived from user-item interactions while simultaneously improving the quality of side information. Our proposed \model\ incorporates three LLM-based strategies for augmenting the interaction graph:
\emph{i)} Reinforcing user-item
interaction edges, \emph{ii)} Enhancing item attribute modeling, and \emph{iii)} Conducting user profiling.
To tackle \textbf{C1} for '\emph{i)}', we devise an LLM-based Bayesian Personalized Ranking (BPR)\cite{rendle2012bpr} sampling algorithm. This algorithm uncover items that users may like or dislike based on textual content from from natural language perspective. These items are then used as positive and negative samples in the BPR training process. It is important to note that LLMs are unable to perform all-item ranking, so the selected items are chosen from a candidate item pool provided by the base recommender for each user.
During the node attribute generation process (corresponding to '\emph{ii)}' and '\emph{iii)}'), we create additional attributes for each user/item using existing text and interaction history. However, it is important to acknowledge that both the augmented edges and node features can contain noise. To address \textbf{C2}, our denoised data robustification mechanism comes into play by integrating noisy edge pruning and feature MAE~\cite{tian2023heterogeneous} to ensure the quality of the augmented data.

In summary, our contributions can be outlined as follows:  \\\vspace{-0.12in}

\begin{itemize}[leftmargin=*]  
\item The \model\ is the pioneering work that using LLMs for graph augmentation in recommender by augmenting: user-item interaction edges, ii) item node attributes, iii) user node profiles.\\\vspace{-0.12in}


\item The proposed \model\ addresses the scarcity of implicit feedback signals by enabling LLMs to reason explicitly about user-item interaction patterns. Additionally, it resolves the low-quality side information issue through user/item attribute generation and a denoised augmentation robustification mechanism with the noisy feedback pruning and MAE-based feature enhancement. \\\vspace{-0.12in}
\\\vspace{-0.12in}


\item Our method has been extensively evaluated on real-world datasets, demonstrating its superiority over state-of-the-art baseline methods. The results highlight the effectiveness of our approach in improving recommendation accuracy and addressing sparsity issues. Furthermore, in-depth analysis and ablation studies provide valuable insights into the impact of our LLM-enhanced data augmentation strategies, further solidifying the model efficacy.
\end{itemize}

%% file: pre.tex
\section{Preliminary}
\label{sec:model}
\textbf{Recommendation with Graph Embedding}.    
Collaborative filtering (CF) learns from sparse implicit feedback $\mathcal{E}^+$,
with the aim of learning collaborative ID-corresponding embeddings $\textbf{E}_u, \textbf{E}_i $ for recommender prediction, given user $u \in \mathcal{U}$ and item  $ i \in \mathcal{I}$. 
Recent advanced recommenders employ GNNs to model complex high-order\cite{tian2023knowledge} u-i relation by taking $\mathcal{E}^+$ as edges of sparse interactive graph.
Therefore, the CF process can be separated into two stages, bipartite graph embedding, and u-i prediction.
Optimizing collaborative graph embeddings $ \textbf{E} = 
\{ \textbf{E}_u, \textbf{E}_i \}$ aims to maximize the posterior estimator with $\mathcal{E}^+$, which is formally presented below: 
\setlength{\abovedisplayskip}{1.0pt} 
\setlength{\belowdisplayskip}{0.5pt}
\begin{align}
    \textbf{E}^* = \argmax\limits_{ \textbf{E} } p(  \textbf{E} | \mathcal{E}^+ ) 
\end{align}
Here, $p(\textbf{E}  |\mathcal{E}^+)$ is to encode as much u-i relation from $\mathcal{E}^+$ into $ \textbf{E}_u, \textbf{E}_i $ as possible for accurate u-i prediction $\hat{y}_{u,i} = \textbf{e}_u \cdot \textbf{e}_{i}$.

\noindent \textbf{Recommendation with Side Information}. However, sparse interactions in $\mathcal{E}^+$ pose a challenge for optimizing the embeddings. 
To handle data sparsity, many efforts introduced side information in form of node features $\textbf{F}$, by taking recommender encoder $f_{\Theta}$ as feature graph.
The learning process of the $f_{\Theta}$ (including $\textbf{E}_u, \textbf{E}_i$ and feature encoder) with side information $\textbf{F}$ is formulated as maximizing the posterior estimator $p(\Theta|\textbf{F}, \mathcal{E}^+)$:
\begin{align}
    \Theta^* = \argmax\limits_{ \Theta } p( \Theta | \textbf{F}, \mathcal{E}^+ ) 
\end{align}
$f_{\Theta}$ will output the final representation $\textbf{h}$ contain both collaborative signals from $\textbf{E}$ and side information from $\textbf{F}$, \ie, $ \textbf{h} = f_{\Theta}(\textbf{f}, \mathcal{E}^+)$. 
\\\vspace{-0.12in}

\noindent \textbf{Recommendation with Data Augmentation}.
Despite significant progress in incorporating side information into recommender, introducing low-quality side information may even undermine the effectiveness of sparse interactions $\mathcal{E}^+$.
To address this, our \model\ focuses on user-item interaction feature graph augmentation, which involves 
LLM-augmented u-i interactive edges  $\mathcal{E}_{\mathcal{A}}$, and LLM-generated node features $\textbf{F}_{\mathcal{A}}$. The optimization target with augmented interaction feature graph is as: 
\begin{align}
    \Theta^* = \argmax\limits_{ \Theta } p( \Theta |  \{ \textbf{F},\textbf{F}_{ \mathcal{A} }  \} ,  \{ \mathcal{E}^+, \mathcal{E}_{ \mathcal{A} }  \}  ) 
\end{align}
The recommender $f_{\Theta}$ input union of original and augmented data, which consist of edges $ \{ \mathcal{E}^+, \mathcal{E}_{ \mathcal{A} }  \} $ and node features  $ \{ \textbf{F},\textbf{F}_{ \mathcal{A} } \}$,
and output quality representation $\textbf{h}$ to predicted preference scores $\hat{y}_{u,i}$ by ranking the likelihood of user $u$ will interact with item $i$. 

%% file: solution.tex
\section{Methodology}
\label{sec:method}


To conduct LLM-based augmentation, in this section, we address these questions:
\textbf{Q1:} How to enable LLMs to predict u-i interactive edges? 
\textbf{Q2:} How to enable LLMs to generate valuable content? 
\textbf{Q3:} How to incorporate augmented contents into original graph contents?
\textbf{Q4:} How to make model robust to the augmented data?  

\begin{figure*}
    \centering
\includegraphics[width=2.1\columnwidth]{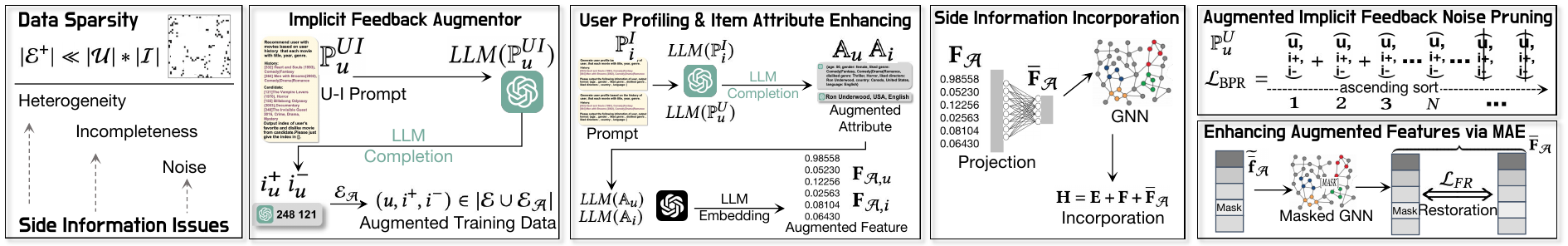}
    \vspace{-0.22 in}
    \caption{
     The \model\ framework: (1) Three types of data augmentation strategies: i) augmenting user-item interactions; ii) enhancing item attributes, and iii) user profiling.
    (2) Augmented training with and denoised data robustification mechanism.
    }
    \vspace{-1.0 em}
    \label{fig:framework}
\end{figure*}

\subsection{LLMs as Implicit Feedback Augmentor (Q1)}
\label{sec:u-i-aug}
\begin{figure}[t]
    \centering
     \includegraphics[width=1\columnwidth]{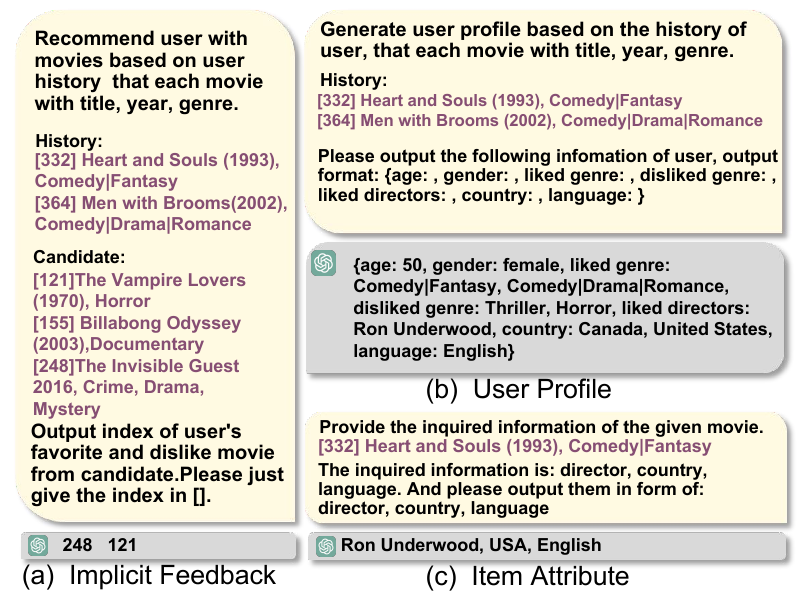}
    \vspace{-0.2in}
    \caption{ 
    {Constructed prompt $\mathbb{P}^{UI}_u, \mathbb{P}^{U}_u, \mathbb{P}^{I}_i$ for LLMs' completion including i) task description, ii) historical interactions, iii) candidates, and iv) output format description.}
}
    \vspace{-1.0 em}
    \label{fig:prompting}
\end{figure}

To directly confront the scarcity of implicit feedback, 
we employ LLM as a knowledge-aware sampler to sample pair-wise~\cite{rendle2012bpr} u-i training data from a natural language perspective. 
This increases potential effective supervision signals and helps gain a better understanding of user preferences by integrating contextual knowledge into the u-i interactions.
Specifically, we feed each user's historical interacted items with side information (\eg, year, genre) and an item candidates pool $\mathcal{C}_u = { \{ i_{u,1}, i_{u,2}, ..., i_{u,|\mathcal{C}_u|} \} }$ into LLM. LLM then is expected to select items that user $u$ might be likely ($i^+_u$) or unlikely ($i^-_u$) to interact with from $\mathcal{C}_u $. 
Here, we introduce $\mathcal{C}_u$ because LLMs can't rank all items. Selecting items from the limited candidate set recommended by the base recommender (\eg, MMSSL~\cite{wei2023multi}, MICRO~\cite{zhang2022latent}), is a practical solution. 
These candidates $\mathcal{C}_u$ are hard samples with high prediction score $\hat{y}_{ui}$ to provide potential, valuable positive samples and hard negative samples.
It is worth noting that we represent each item using textual format instead of ID-corresponding indexes ~\cite{li2023gpt4rec}. This kind of representation offers several advantages: (1) It enables recommender to fully leverage the content in datasets, and (2) It intuitively reflects user preferences. The process of augmenting user-item interactive edges and incorporating it into the training data can be formalized as:  
\begin{equation}
 \label{eq:gan-loss}
    i^+_u, i^-_u = LLM( \mathbb{P}^{UI}_u ); \quad \mathcal{E}_{BPR} = \mathcal{E} \cup \mathcal{E}_{\mathcal{A}}
\end{equation}
where $i^+_u, i^-_u$ are positive and negative samples for BPR selected by LLMs from candidates $\mathcal{C}_u$ for user $u$ based on input prompt $\mathbb{P}^{UI}_u$.
The augmented dataset $\mathcal{E}_{\mathcal{A}}$ comprises pairwise training triplets $(u, i_u^+, i_u^-)$, i.e., $\mathcal{E}_{\mathcal{A}}$ = $ \{ (u, i^+_u, i^-_u) | (u, i^+_u) \in \mathcal{E}^+_{\mathcal{A}}, (u, i^-_u) \in \mathcal{E}^-_{\mathcal{A}} \}$.  
The textual u-i augmentation prompt $\mathbb{P}^{UI}_u$ encompasses different components: i) task description, ii) historical interactions, iii) candidates, and iv) output format description, as illustrated in Fig.~\ref{fig:prompting} (a).


The utilization of LLMs-based sampler in this study 
to some extent alleviate \emph{noise} (\ie, false positive) and \emph{non-interacted items issue
} (\ie, false negative) ~\cite{chen2023revisiting,lee2021bootstrapping} exist in raw implicit feedback. 
In this context, 
(i) \textbf{false positive} are unreliable u-i interactions, which encompass items that were not genuinely intended by the user, such as accidental clicks or instances influenced by popularity bias~~\cite{wang2021denoising}; (ii) \textbf{false negative} represented by non-interacted items, which may not necessarily indicate user dispreference but are conventionally treated as negative samples~~\cite{chen2020efficient}. 
By taking LLMs as implicit feedback augmentor, \model\ enables the acquisition of more meaningful and informative samples by leveraging the remarkable reasoning ability of LLMs with the support of LLMs' knowledge. The specific analysis is supported by theoretical discussion in Sec.~\ref{sec:gradient}.  

\vspace{-0.12in}
\subsection{LLM-based Side Information Augmentation}
\subsubsection{\bf User Profiling \& Item Attribute Enhancing  (Q2)} Leveraging knowledge base and reasoning abilities of LLMs, we propose to summarize user profiles by utilizing users' historical interactions and item information to overcome limitation of privacy. Additionally, the LLM-based item attributes generation aims to produce space-unified, and informative item attributes. Our LLM-based side information augmentation paradigm consists of two steps: 

\begin{itemize}[leftmargin=*]
\item i) \textbf{User/Item Information Refinement}. 
Using prompts derived from the dataset's interactions and side information, we enable LLM to generate user and item attributes that were not originally part of the dataset. 
Specific examples are shown in Fig.~\ref{fig:prompting}(b)(c).

\item ii) \textbf{LLM-enhanced Semantic Embedding}. 
The augmented user and item information will be encoded as features and used as input for the recommender. 
Using LLM as an encoder offers efficient and state-of-the-art language understanding, enabling profiling user interaction preferences and debiasing item attributes. 

\end{itemize}
\noindent Formally, the LLM-based side information augmentation is as:
\begin{align}
\begin{cases} user:  &  \mathbb{A}_{u} = LLM( \mathbb{P}^{U}_u) \quad \longrightarrow \quad \textbf{f}_{ \mathcal{A},u} = LLM(\mathbb{A}_{u}) \\   
    item: &   \mathbb{A}_{i} = LLM( \mathbb{P}^{I}_i) \quad \longrightarrow \quad \textbf{f}_{\mathcal{A}, i} = LLM(\mathbb{A}_{i})  
\end{cases}
\label{eq:att-gene}
\end{align}
\noindent where $ \textbf{f}_{ \mathcal{A}, u}, \textbf{f}_{\mathcal{A}, i, }  \in \mathbb{R}^{ d_{LLM}} $ are LLM-augmented user/item features with LLM's hidden dimension $ d_{LLM} $. 
The textual prompts $\mathbb{P}^{U}_u$ and $\mathbb{P}^{I}_i$ are used for attribute refinement for user $u$ and item $i$, respectively.
$\mathbb{A}_{u}$ and $\mathbb{A}_{i}$ represent generated textual attributes that to be encoded as features $ \textbf{F}_{ \mathcal{A}, u}, \textbf{F}_{\mathcal{A}, i}  $ using the embedding capability of $LLM(\cdot)$.

\subsubsection{ \bf Side Information Incorporation  (Q3)}
\label{sec:incorporation}
After obtaining the augmented side information for user/item, an effective incorporation method is necessary. 
\model\ includes a standard procedure: (1) Augmented Semantic Projection, (2) Collaborative Context Injection, and (3) Feature Incorporation. Let's delve into each:

\begin{itemize}[leftmargin=*]

\item \textbf{Augmented Semantic Projection}. 
Linear layers with dropout are employed to not only reduce the dimensionality of LLM-enhanced semantic features but also map such augmented features into their own space~\cite{wei2023multir}. 
This process can be represented as $\overline{\textbf{F}}_{\mathcal{A}} = \text{Linear}(\textbf{F}_{\mathcal{A}})$, where $\textbf{f}_{\mathcal{A}} \in \mathbb{R}^{1 \times d_{LLM}}$ is the input feature and $\overline{\textbf{f}}_{\mathcal{A}} \in \mathbb{R}^{1 \times d}$ is the output feature after projection.  

\item \textbf{Collaborative Context Injection}. 
\label{sec:side-incorporation}
To inject high-order~\cite{wang2019neural} collaborative connectivity into augmented features $\overline{\textbf{f}}_{\mathcal{A}, u}$ and $\overline{\textbf{f}}_{\mathcal{A}, i}$, \model\ employs light weight GNNs~\cite{he2020lightgcn} as the encoder. 

\item \textbf{Semantic Feature Incorporation}. 
Instead of taking augmented features $\overline{\textbf{F}}_{\mathcal{A}}$ as initialization of learnable vectors of recommender $f_{\Theta}$, 
we opt to treat $\overline{\textbf{F}}_{\mathcal{A}}$ as additional compositions added to the ID-corresponding embeddings ($\textbf{e}_u$, $\textbf{e}_i$). 
This allows flexibly adjust the influence of LLM-augmented features using scale factors and normalization.
Formally, the $\overline{\textbf{F}}_{\mathcal{A}}$'s incorporation is presented as:
\end{itemize}
\setlength{\abovedisplayskip}{1.0pt} 
\setlength{\belowdisplayskip}{0.5pt}  
\begin{align}
    \textbf{h}_u = \textbf{e}_{u} + \omega_1  \cdot \sum_{ k \in \mathcal{M} \cup \mathbb{A}_u }^{|\mathcal{M}| + |\mathbb{A}_u|}{   \frac{ \overline{\textbf{f}}_{u}^{k}}{\| \overline{\textbf{f}}_u^k\|_2}}; ~~~~
    \textbf{h}_i = \textbf{e}_{i} + \omega_1  \cdot \sum_{ k \in \mathcal{M} \cup \mathbb{A}_i }^{|\mathcal{M}| + |\mathbb{A}_i|}{  \frac{ \overline{\textbf{f}}_{i}^{k}}{\| \overline{\textbf{f}}_i^k\|_2}} \nonumber  
\end{align} 
\noindent The final prediction representations $\textbf{h}_u$ and $\textbf{h}_i$, are in $\mathbb{R}^{1 \times d}$. User profiles are $\mathbb{A}_u$, debiased item attributes are $\mathbb{A}_i$, and original multi-modal side information is $\mathcal{M}$. 
The specific type of feature is $\textbf{f}^k$.
We adjust feature vectors using the aggregation weight $\omega_1$ and $L_2$ normalization to mitigate distribution gaps \citep{fu2020magnn},  ensurring the effectiveness of additional features within the recommender encoder. 
\vspace{ -0.3 in} 



\subsection{Training with Denoised Robustification (Q4)} 
\label{sec:quality}
In this section, we outline how \model\ integrate augmented data into the optimization.  
We also introduce two quality constraint mechanisms for augmented edges and node features: i) Noisy user-item interaction pruning, and ii) MAE-based feature enhancement. 
\vspace{ -0.25 in}

\subsubsection{ \bf Augmented Optimization with Noise Pruning}
\label{sec:loss-prune} 
We train our recommender using the union set $\mathcal{E} \cup \mathcal{E}_{\mathcal{A}}$, which includes the original training set $\mathcal{E}$ and the LLM-augmented set $\mathcal{E}_{\mathcal{A}}$. The objective is to optimize the BPR $\mathcal{L}_\text{BPR}$ loss with increased supervisory signals $\mathcal{E} \cup \mathcal{E}_{\mathcal{A}}$, aiming to enhance the recommender's performance by leveraging the incorporated LLM-enhanced user preference:
\begin{align}
\label{eq:bpr}
\mathcal{L}_\text{BPR} & = \sum_{(u,i^+,i^-)}^{|\mathcal{E} \cup \mathcal{E}_{ \mathcal{A} }  |} - \log \left( \sigma (\hat{y}_{u,i^+} - \hat{y}_{u,i^-})\right) + \omega_2\cdot \|\Theta\|^2  \\
& \mathcal{E}_{ \mathcal{A} }  \subseteq  \{ LLM( \mathbb{P}_{u}) | u \in \mathcal{U} \}, \quad |\mathcal{E}_{ \mathcal{A} }| = \omega_3*B
 \nonumber
\end{align}
\noindent The training triplet $(u,i^+,i^-)$ is selected from the union training set $\mathcal{E} \cup \mathcal{E}_{\mathcal{A}}$. 
The predicted scores of positive-negative sample pairs are obtained through inner products of final representation $\textbf{h}$, \ie, $\hat{y}_{u,i^+} = \textbf{h}_u \cdot \textbf{h}_{i+}, \hat{y}_{u,i^-} = \textbf{h}_u \cdot \textbf{h}_{i-}$.
The augmented dataset $\mathcal{E}_{\mathcal{A}}$ is a subset of the overall LLM-generated data $ \{ LLM( \mathbb{P}_{u}) | u \in \mathcal{U} \}$, obtained by sampling. 
This is because excessive inclusion of pseudo label may lead to a degradation in result accuracy. 
The number of samples $| \mathcal{E}_{ \mathcal{A} } |$ is controlled by the batch size $B$ and a rate $\omega_3$.  Weight-decay regularization $|\Theta|^2$ weighted by $\omega_2$, mitigates overfitting.
$\sigma(\cdot)$ is activation function sigmoid to introduce non-linearity. 

\noindent \textbf{Noise Pruning}. To enhance the effectiveness of augmented data, we prune out unreliable u-i interaction noise. Technically, the largest values before minus are discarded after sorting each iteration. This helps prioritize and emphasize relevant supervisory signals while mitigating the influence of noise. Formally, the objective $\mathcal{L}_\text{BPR}$ in Eq.~\ref{eq:bpr} with noise pruning can be rewritten as follows:
\begin{equation}  
\small
   \sum_{(u,i^+,i^-)}^{ (1-\omega_4)*|\mathcal{E} \cup \mathcal{E}_{ \mathcal{A}}| } - SortAscend\left( \log ( \sigma (\hat{y}_{u,i^+} - \hat{y}_{u,i^-}) ) \right)[0:N] + \omega_2\cdot \|\Theta\|^2 
    \label{eq:prune-loss}
\end{equation}
The function $SortAscend(\cdot)[0:N]$ sorts values and selects the top-N. The retained number $N$ is calculated by $N = (1-\omega_4) \cdot |\mathcal{E} \cup \mathcal{E}_{\mathcal{A}}|$, where $\omega_4$ is a rate. This approach allows for controlled pruning of loss samples, emphasizing relevant signals while reducing noise. 
This can avoid the impact of unreliable gradient backpropagation, thus making optimization more stable and effective.

\subsubsection{ \bf Enhancing Augmented Semantic Features via MAE}
To mitigate the impact of noisy augmented features, we employ the Masked Autoencoders (MAE) for feature enhancement ~\cite{he2022masked}. 
Specifically, the masking technique is to reduce the model's sensitivity to features, and subsequently, the feature encoders are strengthened through reconstruction objectives.
Formally, we select a subset of nodes $\widetilde{\mathcal{V}} \subset \mathcal{V}$ and mask their features using a mask token [MASK], denoted as $\textbf{f}_{[MASK]} $ (\eg, a learnable vector or mean pooling).
The mask operation can be formulated as follows:
\begin{equation}
 \label{eq:gan-loss}
    \widetilde{ \overline{\textbf{f}}}_{ \mathcal{A} } =  
    \begin{cases}
    \textbf{f}_{[MASK]} & v \in \widetilde{\mathcal{V}} \\
    \overline{\textbf{f}}_{ \mathcal{A} } &   v \notin \widetilde{\mathcal{V}}
    \end{cases}
\end{equation} 
\noindent The augmented feature after the mask operation is denoted as $\widetilde{\overline{\textbf{f}}}_{\mathcal{A}}$. It is substituted as mask token $\textbf{f}_{[MASK]}$ if the node is selected ($\widetilde{\mathcal{V}} \subset \mathcal{V}$), otherwise, it corresponds to the original augmented feature $\overline{\textbf{f}}_{\mathcal{A}}$. To strengthen the feature encoder, we introduce the feature restoration loss $\mathcal{L}_{FR}$ by comparing the masked attribute matrix $\widetilde{\overline{\textbf{f}}}_{\mathcal{A}, i}$ with the original augmented feature matrix $\overline{\textbf{f}}_{\mathcal{A}}$, with a scaling factor $\gamma$. The restoration loss function $\mathcal{L}_{FR}$ is as follows:
\begin{equation}
    \mathcal{L}_{FR} =  \frac{1}{| \widetilde{ \mathcal{V} }|} \sum\limits_{ v \in \widetilde{\mathcal{V}}   } (  1 - \frac{ \widetilde{ \overline{\textbf{f}}}_{ \mathcal{A}}  \cdot  \overline{\textbf{f}}_{ \mathcal{A}}  }{ \| \widetilde{ \overline{\textbf{f}}}_{ \mathcal{A}} \| \cdot  
\| \overline{\textbf{f}}_{ \mathcal{A}}   \| } )^{\gamma}
\end{equation}
\noindent The final optimization objective is the weighted sum of the noise-pruned BPR loss $\mathcal{L}_\text{BPR}$ and the feature restoration (FR) loss $\mathcal{L}_{FR}$.

\subsection{In-Depth Analysis of our \model}

\subsubsection{ \bf LLM-based Augmentation Facilitates Optimization}
\label{sec:gradient}
\begin{figure}[t]
    \centering
     \includegraphics[width=1\columnwidth]{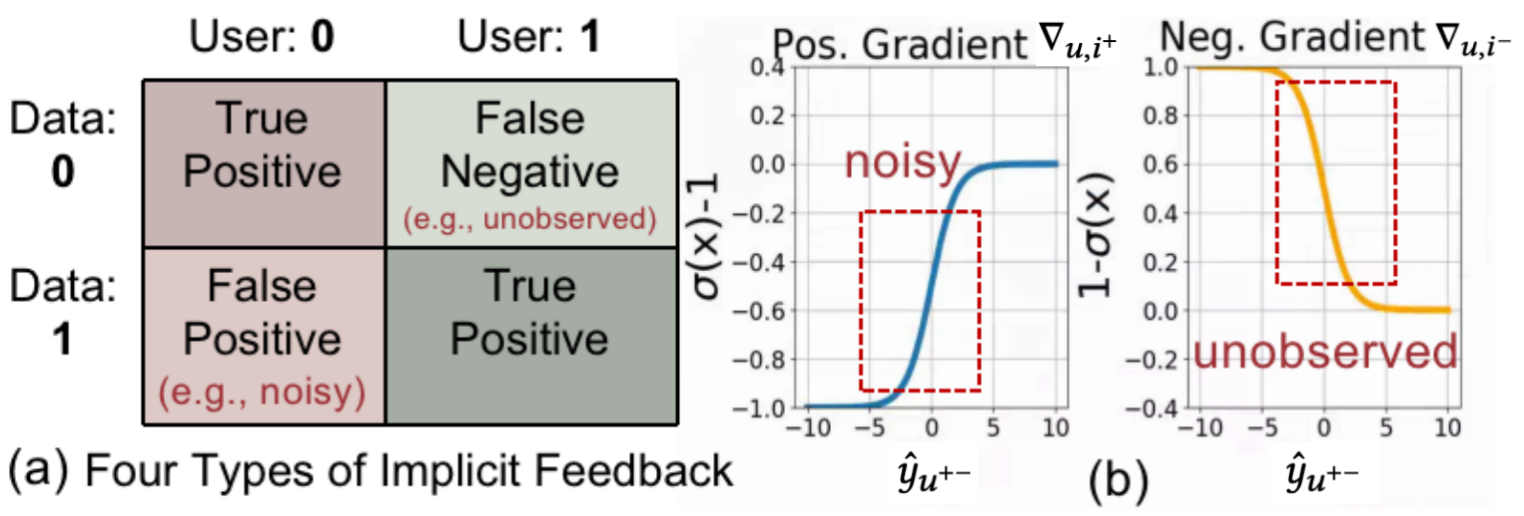}
    \vspace{-0.25in}
    \caption{ (a) Implicit feedback encompasses both false positive and false negative samples. (b) The gradient $\nabla$ of the BPR loss for positive $\hat{y}_{u,i^{+}}$ and negative $\hat{y}_{u,i^{-}}$ scores, despite having a large magnitude, can have an incorrect direction that notably impacts the robustness and effectiveness of training.
  }
    \vspace{-1.5 em}
    \label{fig:gradient}
\end{figure}

This section highlights challenges addressed by LLM-based augmentation in recommender systems. False negatives (non-interacted interactions) and false positives (noise) as in Fig.~\ref{fig:gradient} (a) can affect data quality and result accuracy \citep{chen2020efficient,wang2021denoising}. Non-interacted items do not necessarily imply dislike \citep{chen2020efficient}, and interacted one may fail to reflect real user preferences due to accidental clicks or misleading titles, etc.  
Mixing of unreliable data with true user preference poses a challenge in build accurate recommender. Identifying and utilizing reliable examples is key to optimizing the recommender \citep{chen2023revisiting}. 







In theory, non-interacted and noisy interactions are used for negative $\hat{y}_{u,i^-}$ and positive $\hat{y}_{u,i^+}$ scores, respectively. 
However, their optimization directions oppose the true direction with large magnitudes, \ie, the model optimizes significantly in the wrong directions (as in Fig.~\ref{fig:gradient} (b)), resulting in sensitive suboptimal results. \\
\textbf{Details.} By computing the derivatives of the $ \mathcal{L}_{BPR} $ ( Eq.~\ref{eq:gradient}), we obtain positive gradients $\nabla_{u,i^+}$ = $1 - \sigma( \hat{y}_{u^{+-}} )$ and negative gradients $\nabla_{u,i^{-}}$ = $\sigma( \hat{y}_{u^{+-}} ) - 1$, where $ \hat{y}_{u^{+-}} = \hat{y}_{u,i^{+}} - \hat{y}_{u,i^{-}}$. 
Fig.~\ref{fig:gradient} (b) illustrates these gradients and unveils some observations. Noisy interactions, although treated as positives, often have small values $\hat{y}_{u,i^{+}}$ as false positives, resulting in large gradients $\nabla_{u,i^{+}}$. Conversely, unobserved items, treated as negatives, tend to have relatively large values $\hat{y}_{u,i^{-}}$ as false negatives, leading to small $\hat{y}_{u^{+-}}$ and large gradients $\nabla_{u,i^{-}}$. 

\begin{equation}
\small
 \tag{10} 
\left\{
    \begin{aligned}
     \nabla_{u,i^{+}} & = \frac{ \partial \mathcal{L}_{BPR} }{ \partial \hat{y}_{u,i^{+} } }  = - \frac{ \partial \log \sigma( \hat{y}_{u^{+-}} ) }{ \partial \sigma ( \hat{y}_{u^{+-}} ) } \frac{ \partial \sigma ( \hat{y}_{u^{+-}} ) }{ \partial \hat{y}_{u,i^{+}} } \\[-10pt]    
     & =  - \frac{ 1 }{ \sigma( \hat{y}_{u^{+-}} ) } \cdot \sigma( \hat{y}_{u^{+-}} ) \cdot ( 1 - \sigma( \hat{y}_{u^{+-}} ) ) \cdot 1 = \sigma( \overbrace{ \hat{y}_{u,i^{+}} }\limits^{noisy}   - \hat{y}_{u,i^{-}} ) - 1  \\
     \nabla_{u,i^{-}} & = \frac{ \partial \mathcal{L}_{BPR} }{ \partial \hat{y}_{u,i^{-}} }  = - \frac{ \partial \log \sigma( \hat{y}_{u^{+-}} ) }{ \partial \sigma ( \hat{y}_{u^{+-}} ) } \frac{ \partial \sigma ( \hat{y}_{u^{+-}} ) }{ \partial \hat{y}_{u,i^{-}} } \\[-12pt]     
     & = \frac{ 1 }{ \sigma( \hat{y}_{u^{+-}} ) } \cdot \sigma( \hat{y}_{u^{+-}} ) \cdot ( 1 - \sigma( \hat{y}_{u^{+-}} ) ) \cdot 1 = 1 - \sigma( \hat{y}_{u,i^{+}} - \overbrace{\hat{y}_{u,i^{-}}}\limits^{unobserved}  )
\end{aligned}
\right.
\label{eq:gradient}
\end{equation}

\noindent \textbf{Conclusion.} Wrong samples possess incorrect directions but are influential. LLM-based augmentation uses the natural language space to assist the ID vector space to provide a comprehensive reflection of user preferences. 
With real-world knowledge, \model\ gets quality samples, reducing the impact of noisy and unobserved implicit feedback, improving accuracy, and speeding up convergence.

\vspace{-0.05in}
\subsubsection{ \bf Time Complexity.} 
We analyze the time complexity. The projection of augmented semantic features has a time complexity of $\mathcal{O}(|\mathcal{U} \cup \mathcal{I}| \times d_{LLM} \times d)$. The GNN encoder for graph-based collaborative context learning takes $\mathcal{O}(L \times |\mathcal{E}^+| \times d)$ time. The BPR loss function computation has a time complexity of $\mathcal{O}(d \times |\mathcal{E} \cup \mathcal{E_{\mathcal{A}}}|)$, while the feature reconstruction loss has a time complexity of $\mathcal{O}(d \times |\widetilde{\mathcal{V}}|)$, where $|\widetilde{\mathcal{V}}|$ represents the count of masked nodes.

%% file: eval.tex
\section{Evaluation}
\label{sec:eval}

\begin{table}[]
\caption{Statistics of the Original and Augmented Datasets}
\vspace{-0.15 in}
\setlength{\tabcolsep}{1.0mm}
\small
\begin{tabular}{|cc|ccccccc|}
\hline
\multicolumn{2}{|c|}{Dataset}                                        & \multicolumn{3}{c|}{Netflix}                                          & \multicolumn{1}{c|}{}   & \multicolumn{3}{c|}{MovieLens}                                               \\ \hline  \hline
\multicolumn{1}{|c|}{\multirow{3}{*}{Graph}} & \multirow{2}{*}{Ori.} & \# U                        & \# I     & \multicolumn{1}{c|}{\# E}    & \multicolumn{1}{c|}{}   & \# U                          & \# I                 & \# E                  \\
\multicolumn{1}{|c|}{}                       &                       & 13187                       & 17366    & \multicolumn{1}{c|}{68933}   & \multicolumn{1}{c|}{}   & 12495                         & 10322                & 57960                \\ \cline{2-5} \cline{7-9} 
\multicolumn{1}{|c|}{}                       & \textbf{Aug.}                  & \textbf{\# E:}                       & \textbf{26374}          & \multicolumn{1}{c|}{}        & \multicolumn{1}{c|}{}   & \textbf{\# E: }                        &   \textbf{24990}                   &                       \\ \hline
\multicolumn{2}{|c|}{Ori. Sparsity}                                  & \multicolumn{3}{c|}{99.970\%}                                         & \multicolumn{1}{c|}{}   & \multicolumn{3}{c|}{99.915\%}                                                \\ \hline
\multicolumn{1}{|c|}{\multirow{3}{*}{Att.}}  & Ori.                  & \multicolumn{1}{c|}{U: None} & \multicolumn{2}{c|}{I: year, title}     & \multicolumn{1}{c|}{}   & \multicolumn{1}{c|}{U: None}  & \multicolumn{2}{c|}{I: title, year, genre}   \\ \cline{2-9} 
\multicolumn{1}{|c|}{}                       & \multirow{2}{*}{\textbf{Aug.}} & \textbf{U[1536]}:                       & \multicolumn{6}{l|}{\begin{tabular}[c]{@{}l@{}} \textbf{age, gender, liked genre, disliked genre,} \\ \textbf{liked directors, country, and language} \end{tabular}} \\ \cline{3-9} 
\multicolumn{1}{|c|}{}                       &                       & \textbf{I[1536]}:                       & \multicolumn{6}{l|}{ \textbf{director, country, language}}                                                                                                 \\ \hline
\multicolumn{2}{|c|}{Modality}                                       & \multicolumn{3}{c|}{Textual[768], Visiual [512]}                       & \multicolumn{1}{c|}{}   & \multicolumn{3}{c|}{Textual [768], Visiual [512]}                              \\ \hline
\end{tabular}
\begin{tablenotes}
\scriptsize
\vspace{1.0 mm}    
\item{ * Att. represents attribute, Ori. represents original, and Aug. represents augmentation. Number in [] represents the feature dimensionality. }  
\end{tablenotes}
\vspace{-0.2in}
\label{tab:dataset}
\end{table}

To evaluate the performance of \model, we conduct experiments, aiming to address the following research questions:

\begin{itemize}[leftmargin=*]
\item \textbf{RQ1}: How does our LLM-enhanced recommender perform compared to the current state-of-the-art baselines?

\item \textbf{RQ2}: What is the impact of key components on the performance?

\item \textbf{RQ3}: How sensitive is the model to different parameters?

\item \textbf{RQ3}: Are the data augmentation strategies in our \model\ applicable across different recommendation models?

\item \textbf{RQ5}: What is the computational cost associated with our devised LLM-based data augmentation schemes?
\end{itemize}

\begin{table*}[t]
\small
\caption{\textbf{Performance comparison on different datasets in terms of \emph{Recall}@10/20/50, and \emph{NDCG}@10/20/50, and \emph{Precision}@20.}}
\vspace{-0.08in}
\setlength{\tabcolsep}{1.7mm}
\begin{tabular}{cccccccccccccccc}
\hline
\multirow{2}{*}{Baseline}   & \multicolumn{6}{c}{Netflix}             &      &  & \multicolumn{6}{c}{MovieLens}           &      \\ \cline{2-8} \cline{10-16} 
                            & R@10 & N@10 & R@20 & N@20 & R@50 & N@50 & P@20 &  & R@10 & N@10 & R@20 & N@20 & R@50 & N@50 & P@20 \\ \hline \hline 
\multicolumn{16}{c}{General Collaborative Filtering Methods} \\ \hline 
MF-BPR                      & 0.0282     & 0.0140     & 0.0542     & 0.0205     & 0.0932     &  0.0281    & 0.0027 &  & 0.1890     & 0.0815     & 0.2564     & 0.0985     & 0.3442     & 0.1161     & 0.0128        \\
NGCF                        & 0.0347     & 0.0161     & 0.0699     & 0.0235     & 0.1092     & 0.0336     & 0.0032    &  & 0.2084     & 0.0886     & 0.2926     & 0.1100     & 0.4262     & 0.1362     & 0.0146    \\
LightGCN                    & 0.0352     & 0.0160     & 0.0701     & 0.0238     & 0.1125     & 0.0339     & 0.0032      &  & 0.1994     & 0.0837     & 0.2660     & 0.1005     & 0.3692     & 0.1209     & 0.0133    \\ \hline
\multicolumn{16}{c}{Recommenders with Side Information}                \\ \hline
VBPR                        & 0.0325     & 0.0142     & 0.0553     & 0.0199     & 0.1024     & 0.0291     & 0.0028     &  & 0.2144     & 0.0929     & 0.2980     & 0.1142     & 0.4076     & 0.1361     & 0.0149     \\
MMGCN                       & 0.0363     & 0.0174     & 0.0699     & 0.0249     & 0.1164     & 0.0342  &   0.0033    &  & 0.2314     & 0.1097     & 0.2856     & 0.1233     & 0.4282     & 0.1514     & 0.0147       \\ 
GRCN                        & 0.0379     & 0.0192     & 0.0706     & 0.0257     & 0.1148     & 0.0358     & 0.0035      &  & 0.2384     & 0.1040     & 0.3130     & 0.1236     & 0.4532     & 0.1516     & 0.0150       \\ \hline

\multicolumn{16}{c}{ Data Augmentation Methods }                \\ \hline
LATTICE                     & 0.0433     & 0.0181     & 0.0737     & 0.0259     & 0.1301     & 0.0370  &   0.0036      &  & 0.2116     & 0.0955     & 0.3454     & 0.1268     & 0.4667     & 0.1479    & 0.0167      \\

MICRO                       & \underline{0.0466}     & 0.0196     & \underline{0.0764}     & 0.0271     & \underline{0.1306}     & 0.0378     & \underline{0.0038}     &  & 0.2150     & \underline{0.1131}     & \underline{0.3461}     & \underline{0.1468}     & \underline{0.4898}     & \underline{0.1743}    & \underline{0.0175} \\  \hline

\multicolumn{16}{c}{ Self-supervised Methods}  \\ \hline
CLCRec                      & 0.0428     & 0.0217     & 0.0607     & 0.0262     & 0.0981     & 0.0335     & 0.0030   &  & 0.2266     & 0.0971     & 0.3164     & 0.1198     & 0.4488     & 0.1459     & 0.0158     \\ 

MMSSL                       & 0.0455     & \underline{0.0224}     & 0.0743     & \underline{0.0287}     & 0.1257     & \underline{0.0383}  & 0.0037       &  & \underline{0.2482}     & 0.1113     & 0.3354     & 0.1310     & 0.4814     & 0.1616     & 0.0170     \\  \hline

\model\                      & \textbf{0.0531}    & \textbf{0.0272}     & \textbf{0.0829}     & \textbf{0.0347}     & \textbf{0.1382}     & \textbf{0.0456}     & \textbf{0.0041}      &  & \textbf{0.2603}     & \textbf{0.1250}     & \textbf{0.3643}     & \textbf{0.1628}     & \textbf{0.5281}     & \textbf{0.1901}    & \textbf{0.0186}    \\

\emph{p}-value                    & 
 \small{2.9$e^{-4}$}     & \small{3.0$e^{-3}$}     & \small{9.4$e^{-5}$}     & \small{1.5$e^{-3}$}     & \small{2.8$e^{-5}$}     & \small{2.2$e^{-3}$}     & \small{3.4$e^{-5}$}     &  & \small{2.8$e^{-5}$}    & 
 \small{1.6$e^{-2}$}     & \small{3.1$e^{-3}$}     & \small{4.1$e^{-4}$}     & \small{1.9$e^{-3}$}     & \small{1.3$e^{-2}$}     & \small{1.8$e^{-3}$}     \\  
Improv.                  & 13.95\%    & 21.43\%     & 8.51\%     & 20.91\%     & 5.82\%     & 19.06\%     & 7.89\%     &  & 4.88\%     & 10.52\%     & 5.26\%     & 10.90\%     & 7.82\%     & 9.06\%    & 6.29\%     \\  \hline
\end{tabular}
\label{tab:overall_performance}
\vspace{-0.01in}
\end{table*}

\vspace{-0.1in}
\subsection{Experimental Settings}

\subsubsection{\bf Datasets.} 
We perform experiments on publicly available datasets, \ie, Netflix and MovieLens, which include multi-modal side information. Tab.~\ref{tab:dataset} presents statistical details for both the original and augmented datasets for both user and item domains. \textbf{MovieLens}. We utilize the MovieLens dataset derived from ML-10M\footnote{https://files.grouplens.org/datasets/movielens/ml-10m-README.html}. Side information includes movie title, year, and genre in textual format. Visual content consists of movie posters obtained through web crawling by ourselves. \textbf{Netflix}. We collected its multi-model side information through web crawling. The implicit feedback and basic attribute are sourced from the Netflix Prize Data\footnote{https://www.kaggle.com/datasets/netflix-inc/netflix-prize-data} on Kaggle. For both datasets, CLIP-ViT\cite{radford2021learning} is utilized to encode visual features. \\\vspace{-0.16in}





\noindent \textbf{LLM-based Data Augmentation}. The study employs the OpenAI package, accessed through LLMs' APIs, for augmentation. The OpenAI Platform documentation provides details\footnote{https://platform.openai.com/docs/api-reference}. Augmented implicit feedback is generated using the "gpt-3.5-turbo-0613" chat completion model. Item attributes such as directors, country, and language are gathered using the same model. User profiling, based on the "gpt-3.5-turbo-16k" model, includes age, gender, preferred genre, disliked genre, preferred directors, country, and language. Embedding is performed using the "text-embedding-ada-002" model. The approximate cost of augmentation strategies on two datasets is 15.65 USD, 20.40 USD, and 3.12 USD, respectively.

\vspace{-0.05in}
\subsubsection{\bf Implementation Details}
The experiments are conducted on a 24 GB Nvidia RTX 3090 GPU using PyTorch\cite{paszke2019pytorch} for code implementation. The AdamW optimizer\cite{loshchilov2017decoupled} is used for training, with different learning rate ranges of [$5e^{-5}$, $1e^{-3}$] and [$2.5e^{-4}$, $9.5e^{-4}$] for Netflix and MovieLens, respectively. Regarding the parameters of the LLMs, we choose the \emph{temperature} from larger values \{0.0, 0.6, 0.8, 1 \} to control the randomness of the generated text. The value of \emph{top-p} is selected from smaller values \{0.0, 0.1, 0.4, 1\} to encourage probable choices. The \emph{stream} is set to false to ensure the completeness of responses. For more details on the parameter analysis, please refer to Section \ref{sec:parameter}. To maintain fairness, both our method and the baselines employ a unified embedding size of 64.

\subsubsection{\bf Evaluation Protocols}
We evaluate our approach in the top-K item recommendation task using three common metrics: Recall (R@k), Normalized Discounted Cumulative Gain (N@k), and Precision (P@k). To avoid potential biases from test sampling, we employ the all-ranking strategy\cite{wei2021hierarchical, wei2020graph}. We report averaged results from five independent runs, setting K to 10, 20, and 50 (reasonable for all-ranking). Statistical significance analysis is conducted by calculating $p$-values against the best-performing baseline.

\subsubsection{\bf Baseline Description}
Four distinct groups of baseline methods for thorough comparison. i) General CF Methods: \textbf{MF-BPR}~\cite{rendle2012bpr}, \textbf{NGCF}~\cite{wang2019neural} and \textbf{LightGCN}~\cite{he2020lightgcn}. ii) Methods with Side Information: \textbf{VBPR}~\cite{he2016vbpr}, \textbf{MMGCN}~\cite{wei2019mmgcn} and \textbf{GRCN}~\cite{wei2020graph}. iii) Data Augmentation Methods: \textbf{LATTICE}~\cite{zhang2021mining}. iv) Self-supervised Methods: \textbf{CLCRec}~\cite{wei2021contrastive}, \textbf{MMSSL}~\cite{wei2023multi} and \textbf{MICRO}~\cite{zhang2022latent}. 

\vspace{-0.1in}
\subsection{Performance Comparison (RQ1)}
Tab. \ref{tab:overall_performance} compares our proposed \model\ method with baselines. 
\vspace{-0.05in}
\begin{itemize}[leftmargin=*]

\item \textbf{Overall Model Superior Performance}. 
Our \textit{\model} outperforms the baselines by explicitly augmenting u-i interactive edges and enhancing the quality of side information. 
It is worth mentioning that our model based on LATTICE's \cite{zhang2021mining} encoder, consisting of a ID-corresponding encoder and a feature encoder. This improvement underscores the effectiveness of our framework.  
\\\vspace{-0.12in}

\item \textbf{Effectiveness of Side Information Incorporation}. The integration of side information significantly empowers recommenders. Methods like MMSSL \cite{wei2023multi} and MICRO \cite{zhang2022latent} stand out for their effective utilization of multiple modalities of side information and GNNs.
In contrast, approaches rely on limited content, such as VBPR \cite{he2016vbpr} using only visual features, or CF-based architectures like NGCF \cite{wang2019neural}, without side information, yield significantly diminished results. This highlights the importance of valuable content, as relying solely on ID-corresponding records fails to capture the complete u-i relationships. 

\item \textbf{Inaccurate Augmentation yields Limited Benefits}.
Existing methods, such as LATTICE\cite{zhang2021mining}, MICRO\cite{zhang2022latent} that also utilize side information for data augmentation have shown limited improvements compared to our \model. This can be attributed to two main factors: (1) The augmentation of side information with homogeneous relationships (e.g., i-i or u-u) may introduce noise, which can compromise the precise of user preferences. (2) These methods often not direct augmentation of u-i interaction data.  

\item \textbf{Advantage over SSL Approaches.} Self-supervised models like,
MMSSL\cite{wei2023multi}, MICRO\cite{zhang2022latent}, have shown promising results in addressing sparsity through SSL signals. However, they do not surpass the performance of \model, possibly because their augmented self-supervision signals may not align well with the target task of modeling u-i interactions. In contrast, we explicitly tackle the scarcity of training data by directly establishing BPR triplets.

\end{itemize}

\subsection{Ablation and Effectiveness Analyses (RQ2)}

\begin{table}[]  
\caption{Ablation study on key components (\ie, data augmentation strategies, denoised data robustification mechanisms)}  
\vspace{-1.7 mm}
\setlength{\tabcolsep}{0.20mm}
\small 
\begin{tabular}{c|c|cc|cc|cc|c}
\hline
                                    & Metrics   & R@10   & N@10   & R@20   & N@20   & R@50   & N@50   & P@20   \\ \hline \midrule
\multirow{4}{*}{\rotatebox{90}{Aug.}}       & w/o-u-i   & 0.0477 & 0.0239 & 0.0791 & 0.0317 & 0.1376 & 0.0432 & 0.0037 \\
                                    & w/o-u     & 0.0423 & 0.0196 & 0.0656 & 0.0255 & 0.1192 & 0.0360 & 0.0033 \\
                                    & w/o-u\&i  & 0.0309 & 0.0127 & 0.0602 & 0.0202 & 0.1051 & 0.0289 & 0.0030 \\  \midrule
\multirow{3}{*}{\rotatebox{90}{Q. C.}} & w/o-prune    & 0.0504 & 0.0258 & 0.0786 & 0.0328 & 0.1363 & 0.0447 & 0.0039 \\ 
                                    & w/o-QC & 0.0488 & 0.0244 & 0.0786 & 0.0318 & 0.1279 & 0.0416 & 0.0038 \\ \midrule 
                                    & \model\      & \textbf{0.0531} & \textbf{0.0272} & \textbf{0.0829} & \textbf{0.0347} & \textbf{0.1382} & \textbf{0.0456} & \textbf{0.0041} \\ \hline
\end{tabular}
\begin{tablenotes}
\footnotesize
\item{*} ``Aug'': data augmentation operations; Q. C.: denoised data robustification.
\end{tablenotes} 
\label{tab:ablation}
\vspace{-0.1in}
\end{table}

We conduct an ablation study of our proposed \model\ approach to validate its key components, and present the results in Table \ref{tab:ablation}.

\subsubsection{ \bf Effectiveness of Data Augmentation Strategies.}

\begin{itemize}[leftmargin=*,label=$\bullet$]

\item (1). \emph{w/o}-u-i: Disabling the LLM-augmented implicit feedback $\mathcal{E}_{\mathcal{A}}$ results in a significant decrease. 
This indicates that \model\ increases the potential supervision signals by including contextual knowledge, leading to a better grasp of user preferences. 

\item (2). \emph{w/o}-u: Removing our augmentor for user profiling result in a decrease in performance, indicating that our LLM-enhanced user side information can effectively summarize useful user preference profile using historical interactions and item-end knowledge.

\item (3). \emph{w/o}-{u\&i}: when we remove the augmented side information for both users and items ($\textbf{F}_{ \mathcal{A}, u}, \textbf{F}_{ \mathcal{A}, i, 1} $), lower recommendation accuracy is observed. This finding indicates that the LLM-based augmented side information provides valuable augmented data to the recommender system, assisting in obtaining quality and informative representations.

\end{itemize}

\subsubsection{  \bf Impact of the Denoised Data Robustification.}

\begin{itemize}[leftmargin=*,label=$\bullet$]

\item \emph{w/o-prune}: The removal of noise pruning results in worse performance. This suggests that the process of removing noisy implicit feedback signals helps prevent incorrect gradient descent.

\item \emph{w/o-QC}: The performance suffer when both the limits on implicit feedback and semantic feature quality are simultaneously removed (\emph{i.e.,} \emph{w/o-prune + w/o-MAE}). This indicates the benefits of our denoised data robustification mechanism by integrating noise pruning and semantic feature enhancement. 

\end{itemize}

\subsection{ Hyperparameter Analysis (RQ3)}
\label{sec:parameter}

\subsubsection{ \bf Parameters Affecting Augmented Data Quality. }

\begin{table}[]
\small
\caption{Parameter analysis of temperature $\tau$ and top-p $\rho$.}
\vspace{-0.09in}
\setlength{\tabcolsep}{0.5mm}
\begin{tabular}{c|cccc|cccc}
\hline
Para.    & \multicolumn{4}{c|}{Temperature $\tau$} & \multicolumn{4}{c}{Top-p $\rho$} \\ \hline
Metrics & $\tau$=0   & $\tau$=0.6   & $\tau$=0.8   & $\tau$=1  & $\rho$=0  & $\rho$=0.1 & $\rho$=0.4 & $\rho$=1 \\ \hline
R@10    & 0.0558$\uparrow$ & \textbf{0.0531} & 0.0553$\uparrow$ & 0.0531= & 0.0537$\uparrow$ & \textbf{0.0531} & 0.0520$\downarrow$ & 0.0531=  \\
R@20    & 0.0808$\downarrow$ & \textbf{0.0829} & 0.0813$\downarrow$ & 0.0775$\downarrow$ & 0.0802$\downarrow$ & \textbf{0.0829} & 0.0796$\downarrow$ & 0.0770$\downarrow$  \\
R@50    & 0.1344$\downarrow$ & \textbf{0.1382} & 0.1360$\downarrow$ & 0.1312$\downarrow$ & 0.1360$\downarrow$ & \textbf{0.1382} & 0.1344$\downarrow$ & 0.1333$\downarrow$  \\
\hline
\end{tabular}
\label{tab:para-t-p}
\vspace{-0.05in}
\end{table}

\begin{table}[]
\small
 \caption{ Analysis of key parameter (\ie, \# candidate $|\mathcal{C}|$ ) for LLM \wrt \ implicit feedback augmentation $\mathcal{E}_{\mathcal{A}}$.} 
 \setlength{\tabcolsep}{1.4mm}
 \vspace{-0.12in}
\begin{tabular}{c|ccc|c|ccc}
\hline
Data    & \multicolumn{3}{c|}{Netflix} &  & \multicolumn{3}{c}{MovieLens} \\ \hline
Metrics & $|\mathcal{C}|$=3     & $|\mathcal{C}|$=10     & $|\mathcal{C}|$=30    &  & $|\mathcal{C}|$=3     & $|\mathcal{C}|$=10     & $|\mathcal{C}|$=30     \\ \hline
R@20    & 0.0786 $\downarrow$ & \textbf{0.0829} & 0.0808 $\downarrow$         & & 0.3567 $\downarrow$ & \textbf{0.3643} & 0.3695 $\uparrow$          \\
N@20    & 0.0314 $\downarrow$ & \textbf{0.0347} & 0.0330 $\downarrow$        &  & 0.1603 $\downarrow$ & \textbf{0.1628} & 0.1614 $\downarrow$         \\
P@20    & 0.0039 $\downarrow$ & \textbf{0.0041} & 0.0040 $\downarrow$         &  & 0.0179 $\downarrow$ & \textbf{0.0186} & 0.0182 $\downarrow$         \\ \hline
\end{tabular}
\label{tab:para-c}
\vspace{-0.1in}
\end{table}

\begin{figure}[t]
    \centering    \includegraphics[width=1.01\columnwidth]{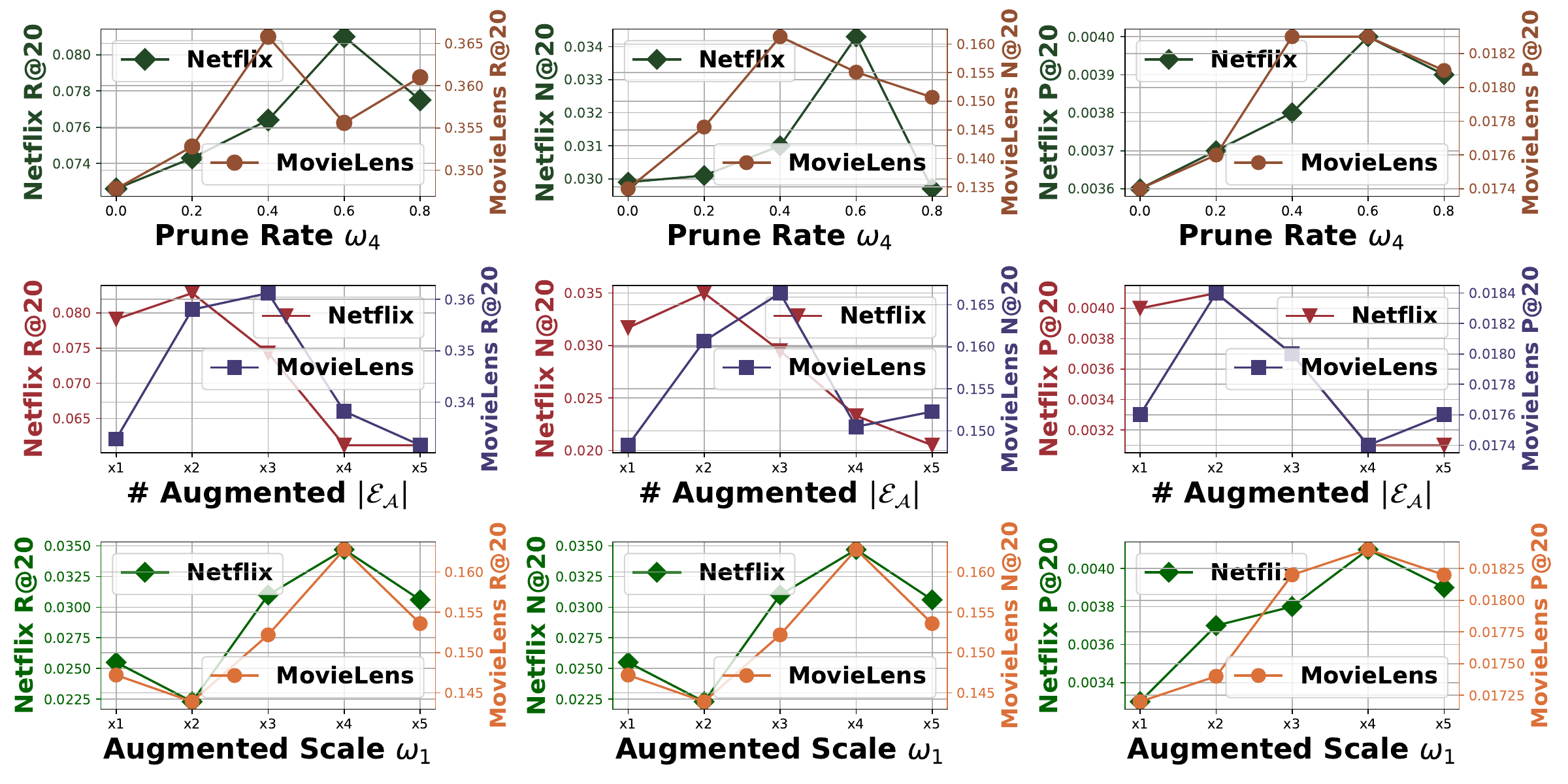}
    \vspace{-0.1in}
    \caption{
     Impact of hyperparameters (\ie, prune rate $\omega_4$, \# augmented BPR training data $|\mathcal{E}_{  \mathcal{A} }|$, and augmented feature incorporate scale $\omega_1$).
     }
    \vspace{-1. em}
    \label{fig:exp-para}
\end{figure}

\begin{itemize}[leftmargin=0.2pt, label=$\bullet$]

\item \emph{Temperature $\tau$ of LLM}: The temperature parameter $\tau$ affects text randomness. Higher values (>1.0) increase diversity and creativity, while lower values (<0.1) result in more focus. We use $\tau$ from $\{0, 0.6, 0.8, 1\}$. As shown in Table \ref{tab:para-t-p}, increasing $\tau$ initially improves most metrics, followed by a decrease.

\item \emph{Top-p $p$ of LLM}: Top-p Sampling\cite{holtzman2019curious} selects tokens based on a threshold determined by the top-p parameter $p$. Lower $p$ values prioritize likely tokens, while higher values encourage diversity. We use $p$ from $\{0, 0.1, 0.4, 1 \}$ and smaller $p$ values tend to yield better results, likely due to avoiding unlisted candidate selection. Higher $\rho$ values cause wasted tokens due to repeated LLM inference.

\item \emph{\# of Candidate $ \mathcal{C}$:} We use $\mathcal{C}$ to limit item candidates for LLM-based recommendation. $\{3, 10, 30\}$ are explored due to cost limitations, and Table \ref{tab:para-c} shows that $\mathcal{C}=10$ yields the best results. Small values limit selection, and large values increase recommendation difficulty.

\item \emph{ Prune Rate $\omega_4 $:} \model\ uses $\omega_4$ to control noise in augmented training data to be pruned. We set $\omega_4$ to \{0.0, 0.2, 0.4, 0.6, 0.8\} on both datasets. As shown in {Fig.~\ref{fig:exp-para} (a)}, $\omega_4=0$ yields the worst result, highlighting the need to constrain noise in implicit feedback. \vspace{ -0.1 pt}

\end{itemize}

\subsubsection{ \bf Sensitivity of Recommenders to the Augmented Data.} 

\begin{itemize}[leftmargin=*] 

\item \emph{\# of Augmented Samples per Batch $|\mathcal{E}_{ \mathcal{A} }|$ }: \model\ uses $\omega_3$ and batch size $B$ to control the number of augmented BPR training data samples per batch. $\omega_3$ is set to \{0.0, 0.1, 0.2, 0.3, 0.4\} on Netflix and \{0.0, 0.2, 0.4, 0.6, 0.8\} on MovieLens. Suboptimal results occur when $\omega_3$ is zero or excessively large. Increasing diversity and randomness can lead to a more robust gradient descent.

\item \emph{ Scale $\omega_2$ for Incorporating Augmented Features}: \model\ uses $\omega_2$ to control feature magnitude, with values set to \{0.0, 0.8, 1.6, 2.4, 3.2\} on Netflix and \{0.0, 0.1, 0.2, 0.3, 0.4\} on MovieLens. Optimal results depend on the data, with suboptimal outcomes occurring when $\omega_2$ is too small or too large, as shown in {Fig.~\ref{fig:exp-para} (c)}.

\end{itemize}

\subsection{Model-agnostic Property (RQ4)}
We conducted model-agnostic experiments on Netflix to validate the applicability of our data augmentation. 
Specifically, we incorporated the augmented implicit feedback $\mathcal{E}_{\mathcal{A}}$ and features $\textbf{F}_{ \mathcal{A}, u}, \textbf{F}_{ \mathcal{A}, i}$ into baselines MICRO, MMSSL, and LATTICE.
As shown in Tab.~\ref{tab:agnostic}, our LLM-based data improved the performance of all models, demonstrating their effectiveness and reusability.
Some results didn't surpass our model, maybe due to: i) the lack of a quality constraint mechanism to regulate the stability and quality of the augmented data, and ii) the absence of modeling collaborative signals in the same vector space, as mentioned in Sec.~\ref{sec:side-incorporation}. 

\begin{table}[]
\vspace{-0.3 pt}
\caption{ Model-agnostic experiment to evaluate the effectiveness of LLM-based data augmentation on different recommender in terms of R@20, N@20, and P@20. } 
\vspace{-0.8 em}
\setlength{\tabcolsep}{1.4mm} 
\small
\begin{tabular}{ccccccccc}
\hline
                           & Method & LATTICE & & MICRO &  & MMSSL \\ \hline
\multirow{3}{*}{\rotatebox{90}{Aug.}} & R@20   & 0.0821 $\uparrow$11.40\%   &  & 0.0835 $\uparrow$9.29\% &  & 0.0833 $\uparrow$11.11\%      \\
                           & N@20   & 0.0287 $\uparrow$10.81\%    &   & 0.0301 $\uparrow$11.07\%    &  &    0.0313 $\uparrow$9.06\%   \\
                           & P@20   & 0.0039 $\uparrow$8.33\%     &   & 0.0041 $\uparrow$7.89\%     &  &    0.0041 $\uparrow$10.81\%   \\ \hline
\end{tabular}
\label{tab:agnostic}
\vspace{-1.4 pt}
\end{table}

\vspace{-0.12in}
\subsection{Cost/Improvement Conversion Rate (RQ5)}
To evaluate the cost-effectiveness of our augmentation strategies, we compute the CIR as presented in Tab.~\ref{tab:cir}. 
The CIR is compared with the ablation of three data augmentation strategies and the best baseline from Tab.~\ref{tab:ablation} and Tab.~\ref{tab:overall_performance}. The cost of the implicit feedback augmentor refers to the price of GPT-3.5 turbo 4K. The cost of side information augmentation includes completion (using GPT-3.5 turbo 4K or 16K) and embedding (using text-embedding-ada-002). We utilize the HuggingFace API tool for tokenizer and counting. The results in Tab.~\ref{tab:cir} show that 'U' (LLM-based user profiling) is the most cost-effective strategy, and the overall investment is worthwhile.

\begin{table}[t]
\small
\caption{Comparison of the cost and improvement rate(CIR) of data augmentation strategies and \model. 'Cost': expenditure of utilizing LLM, 'Imp.': the average improvement rate in R@10/N@10. 'CIR': the ratio of improvement to cost. }
\vspace{-0.15in}
\begin{tabular}{ccccccc}
\hline
       &           & \multicolumn{2}{c}{R@10} &  & \multicolumn{2}{c}{N@10} \\ \cline{3-4} \cline{6-7} 
       & Cost(USD) & Imp.(\%)       & CIR(\%)       &  & Imp.(\%)       & CIR(\%)       \\ \hline
U      & 10.92         & 25.53           & 233.79            &  & 38.78           & 355.13            \\
I      & 1.96          & 2.31           & 117.86            &  &  1.12          & 57.14            \\
U-I    & 8.26          & 11.32           & 137.05            &  & 13.81           & 167.19            \\ \hline
LLMAug & 21.14          & 13.95           &  65.99           &  &  21.43          & 101.37            \\ \hline
\end{tabular}
\vspace{-0.02 in}
\label{tab:cir}
\end{table}

%% file: related.tex
\section{Related Work}
\label{sec:related}

\noindent \textbf{ Content-based Recommendation.} Existing recommenders have explored the use of auxiliary multi-modal side knowledge\cite{liang2023structure,Liangke_Survey}, with methods like VBPR~\cite{he2016vbpr} combine traditional CF with visual features, while MMGCN~\cite{wei2019mmgcn}, GRCN~\cite{wei2020graph} leverage GNNs to capture modality-aware higher-order collaborative signals. Recent approaches MMSSL~\cite{wei2023multi} and MICRO~\cite{zhang2022latent} align modal signals with collaborative signals through contrastive SSL\cite{Liangke_SymCLKG_TKDE}, revealing the informative aspects of modal signals that benefit recommendations. 
However, the data noise, heterogeneity, and incompleteness can introduce bias. To overcome this, \model\ explores LLM-based augmentation to improve the quality of the data. \\\vspace{-0.12in}

\noindent \textbf{Large Language Models (LLMs) for Recommendation}. LLMs have gained attention in recommendation systems, with various efforts to use them for modeling user behavior \citep{wang2023rethinking,ren2023representation,kang2023llms}. LLMs have been employed as an inference model in diverse recommendation tasks, including rating prediction, sequential recommendation, and direct recommendation \citep{bao2023tallrec,zhang2023recommendation,chen2023palr,dai2023uncovering}. 
Some efforts~\cite{tang2023graphgpt,tian2023graph} also tried to utilize LLMs to model structure relations.
However, most previous methods primarily used LLMs as recommenders, abandoning the base model that has been studied for decades. We combine LLM-based data augmentation with classic CF, achieving both result assurance and enhancement concurrently.
\\\vspace{-0.12in}


\noindent \textbf{Data Augmentation for Recommendation}. Extensive research has explored data augmentation in recommendation systems \citep{lee2021bootstrapping,huang2021mixgcf}. Various operations, such as permutation, deletion, swap, insertion, and duplication, have been proposed for sequential recommendation \citep{liu2021augmenting,petrov2022effective}. Commonly used techniques include counterfactual reasoning~\cite{wang2021counterfactual,zhang2021causerec} and contrastive learning \citep{liu2021contrastive}. 
Our \model\ use LLMs as an inference model to augment edge and enhance node features by leveraging consensus knowledge from the large model. 


%% file: conclusion.tex
\section{Conclusion}
\label{sec:conclusoin}

This study focuses on the design of LLM-enhanced models to address the challenges of sparse implicit feedback signals and low-quality side information by profiling user interaction preferences and debiasing item attributes. To ensure the quality of augmented data, a denoised augmentation robustification mechanism is introduced. The effectiveness of \model\ is supported by theoretical analysis and experimental results, demonstrating its superiority over state-of-the-art recommendation techniques on benchmark datasets. Future directions for investigation include integrating causal inference into side information debiasing and exploring counterfactual factors for context-aware user preference.